\pdftrailerid{}
\documentclass{article}

\usepackage{microtype}
\usepackage{hyperref}
\hypersetup{pdfcreator={}, pdfproducer={}}

\usepackage[accepted]{mlsys2025}
\usepackage[T1]{fontenc}
\usepackage[utf8]{inputenc}
\usepackage{amsmath,amssymb,bm}
\usepackage{graphicx}
\usepackage{booktabs,multirow,tabularx,array}
\usepackage{multicol}
\usepackage{capt-of}
\usepackage{xcolor}
\usepackage{tikz}
\usetikzlibrary{arrows.meta,positioning,fit,calc,shapes.geometric}
\usepackage{listings}
\usepackage[capitalize,noabbrev]{cleveref}
\usepackage{placeins}

\newcommand{\maybefig}[2]{\includegraphics[width=#2]{figures/#1}}

\newcommand{\appendixwidebegin}{\end{multicols}\begin{center}}
\newcommand{\appendixwideend}{\end{center}\begin{multicols}{2}}

\newcommand{\mccl}{MCCL}

\DeclareUrlCommand\code{\urlstyle{tt}}
\DeclareUrlCommand\env{\urlstyle{tt}}

\makeatletter
\g@addto@macro{\UrlBreaks}{\do\_\do\.\do\-\do\(\do\,\do\=}
\makeatother

\mlsystitlerunning{MCCL: PyTorch Training Across a Fleet of Macs}

\begin{document}
\twocolumn[
\mlsystitle{MCCL: Training PyTorch Models Across a Fleet of Macs at Skywalker Sound}

\def\isaccepted{}

\begin{mlsysauthorlist}
\ifdefined\isaccepted
  \mlsysauthor{Kevin Sasso}{skywalker}
  \mlsysauthor{Chris Fiegel}{skywalker}
  \mlsysauthor{Francis Aitken}{skywalker}
  \mlsysauthor{Nicolas Tsingos}{skywalker}
  \mlsysauthor{Nicholas Tong}{skywalker}
\else
  \mlsysauthor{Anonymous Authors}{skywalker}
\fi
\end{mlsysauthorlist}

\mlsysaffiliation{skywalker}{Skywalker Sound, Nicasio, California, USA}
\mlsyscorrespondingauthor{Kevin Sasso}{ksasso@skysound.com}
\mlsyskeywords{distributed training, Apple silicon, PyTorch, collective communication}

\vskip 0.3in
\begin{abstract}
Skywalker Sound, a premier audio post-production facility, operates more than 200 M4 Max Mac Studios on a 10\,Gb Ethernet network. When not in use by artists, these machines sit idle, representing massive untapped unified memory and compute capacity. While these fleets are excellent candidates for distributed machine learning alongside existing NVIDIA GPU clusters, PyTorch lacks an accelerated distributed backend for Apple's Metal Performance Shaders (MPS). To bridge this gap, we introduce the Metal Communication Collective Library (MCCL), the first open-source MPS backend for multi-node PyTorch training over switched Ethernet. We evaluate its performance and scalability across 24 nodes, demonstrating that large all-reduce operations reach 98--99.7\% of TCP goodput. Training a 113.5M-parameter Vision Transformer (ViT) scales from 198 to 4,568 samples/s, achieving 95.9\% parallel efficiency. Benchmarked against an NVIDIA A100 GPU, one M4 Max Mac equals roughly 1.7$\times$ the throughput of FP32 matrix multiplication, and 5.5 Macs match an A100 using TF32. Harnessing 100 idle Macs for 12 hours nightly offsets approximately \$200,000 annually in equivalent cloud compute costs. Our results demonstrate that existing Mac infrastructure can be seamlessly repurposed for high performance distributed training without abandoning the standard PyTorch codebase.

\end{abstract}

]

\printAffiliationsAndNotice{}
\section{Introduction}
\label{sec:intro}

\begin{figure*}[t]
\centering
\begin{tikzpicture}[
  node distance=0.55cm and 1.35cm,
  box/.style={draw, rounded corners=2pt, minimum height=0.68cm, align=center,
              font=\footnotesize, inner sep=4pt},
  arr/.style={-{Stealth[length=1.6mm]}, thick},
  note/.style={font=\scriptsize, align=center}]
  \node[box, fill=orange!12, minimum width=2.2cm] (nas) {ZFS storage\\SMB or NFS};
  \node[box, fill=blue!8, right=of nas, minimum width=1.8cm] (loader) {loader\\threads};
  \node[box, fill=blue!8, right=of loader, minimum width=2.0cm] (gpu) {MPS \textit{forward}\\and \textit{backward}};
  \node[box, fill=green!10, right=of gpu, minimum width=1.9cm] (mccl) {MCCL\\all-reduce};
  \node[box, fill=green!10, right=of mccl, minimum width=1.8cm] (peers) {peer Macs\\over 10\,GbE};
  \draw[arr] (nas) -- node[above,note]{audio} (loader);
  \draw[arr] (loader) -- node[above,note]{batch} (gpu);
  \draw[arr] (gpu) -- node[above,note]{buckets} (mccl);
  \draw[arr, <->] (mccl) -- node[above,note]{gradients} (peers);
  \node[note, below=0.22cm of gpu] (memory) {one 128\,GB unified memory pool};
  \draw[gray!70] (loader.south) -- (memory.north west);
  \draw[gray!70] (mccl.south) -- (memory.north east);
  \node[box, fill=gray!8, above=0.45cm of gpu, minimum width=5.2cm] (ray)
       {Ray places one PyTorch worker on each available Mac};
  \draw[arr] (ray.south) -- (gpu.north);
\end{tikzpicture}
\caption{One MCCL training step on a Mac Studio node: audio batches stream from
ZFS storage, \textit{forward} and \textit{backward} run on the MPS GPU in unified memory, and MCCL
averages each finished gradient bucket in place across peers while the GPU keeps
working.}
\label{fig:system}
\end{figure*}
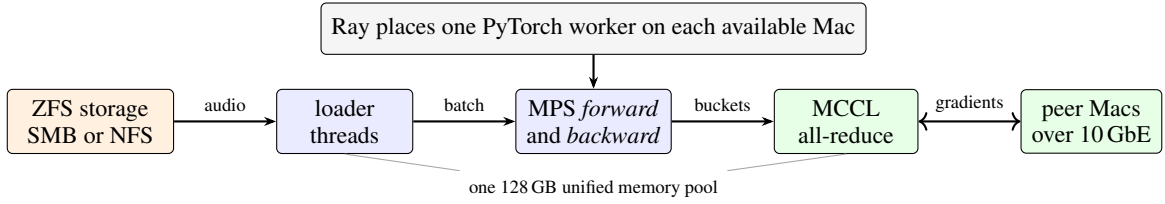

Skywalker Sound, the renowned audio post-production facility behind the iconic soundtracks of \emph{Star
Wars}, \emph{Jurassic Park}, \emph{Indiana Jones} and many others, operates more than 200 Apple M4
Max Mac Studios (40-core GPU, 128\,GB unified memory) on a single 10\,Gb Ethernet network
backed by central ZFS storage. By day, they operate as Digital Audio Workstations to create soundtracks; by night, we estimate that at least 100 of them sit idle for 12 hours, representing 438,000 Mac-hours a year.

Our machine learning team builds custom audio models for tasks such as stem separation, denoising, and
upmixing, and trains them mostly on NVIDIA GPU clusters that are shared across teams and often fully booked. We wanted to use the idle Macs for exploratory prototyping and longer
overnight training runs, with the same PyTorch code we run on our GPU
clusters.
PyTorch, however, has no distributed backend for MPS tensors, and its
CPU backend, Gloo, works only if gradients are copied to CPU memory by
hand, which stalls the GPU on every exchange (\cref{sec:prelim}).
Apple-specific frameworks such as MLX provide their own collectives,
but adopting one would mean maintaining a second codebase.

To fill this gap, we created \mccl{}
(\url{https://github.com/mps-ddp/mccl}), an open-source
\code{ProcessGroup} that registers as \verb|backend="mccl"| and runs
our unmodified models under PyTorch Lightning and Ray. Prior
distributed ML on Apple silicon runs over Thunderbolt. It includes
inference across meshes of four to six
machines~\citep{mlxjaccl,exo2024,mlx_issue3162,genmlx} and, as the
largest training result we are aware of, a 1.7M-parameter model
trained with MLX on 16 Mac minis cabled into a Thunderbolt
ring~\citep{vanderouderaa2025}. The prior multi-Mac PyTorch training we
found used three machines~\citep{jasonappah_cv}. Our evaluation uses 24
Mac Studios on a switched network, with models of up to 113.5M
parameters. \mccl{} lets our team run the same training jobs across all
the accelerator hardware we already own, which matters as GPU capacity,
both on our own clusters and in the cloud, remains contended and
expensive. We expected the 10\,Gb network to be the main obstacle, but
a pipelined TCP ring reached 98--99.7\% of TCP goodput early in
development, and most of our effort went instead into the handoff
between CPU and GPU (\cref{sec:design}). With 100 machines idle each
night, the fleet provides roughly 18 A100 GPUs' worth of training
capacity at little added cost, or about \$200,000 a year in equivalent
cloud compute.\footnote{Assumes 100 Mac Studios idle 12 hours a night
(438,000 Mac-hours a year), one A100 with TF32 matmul matching about
5.5 Macs on our ViT benchmark (\cref{sec:throughput}), and AWS EC2
Capacity Block pricing for p4de.24xlarge of \$20.37/hr for eight A100
80\,GB GPUs, or \$2.546 per GPU-hour
(\url{https://aws.amazon.com/ec2/capacityblocks/pricing/}, accessed
October 20, 2026). Excludes energy and time lost to preemption.}

This paper makes four contributions.
\begin{itemize}
  \item \textbf{MCCL}, an open-source PyTorch \code{ProcessGroup} for
  MPS tensors over switched Ethernet, which runs unmodified
  \code{DistributedDataParallel} code under PyTorch Lightning and Ray
  (\cref{sec:design}).
\item \textbf{An in-place handoff design for unified memory}, in which
CPU threads reduce MPS gradient buckets directly while \textit{backward}
computation continues, and per-bucket Metal shared events enforce the
ordering that prevents races without synchronizing the entire GPU
(\cref{sec:design}).
  \item \textbf{An evaluation on 24 Mac Studios}, to our knowledge the largest reported distributed PyTorch training on Apple silicon, showing
  all-reduce at 98--99.7\% of TCP goodput, about 94\% of communication
  overlapped with \textit{backward} computation, and 95.9\% scaling efficiency
  on a 113.5M-parameter ViT (\cref{sec:eval}).
  \item \textbf{Lessons from scaling past two nodes}, including a
  buffer-reuse race that disappeared whenever we added a
  \code{print()}, a ring-ordering deadlock, and a silent gradient bug
  in PyTorch's MPS reflect-padding kernel that affects STFT losses on
  long audio (\cref{sec:issues}).
\end{itemize}

\section{Preliminaries: Multi-Node PyTorch on Unified Memory}
\label{sec:prelim}

\paragraph{What a \code{ProcessGroup} does and why overlap matters.}
In data-parallel training, every worker holds a full copy of the model, computes
gradients on its slice of the batch, and averages them with other workers before the
optimizer step. The averaging goes through a \code{ProcessGroup}~\citep{ptdistdocs}, the
plugin selected by \code{init_process_group(backend=...)}, which exposes asynchronous
collectives (\emph{all-reduce}, \emph{broadcast}, \emph{allgather},
\emph{reduce-scatter}, \emph{barrier}). PyTorch does not wait for \textit{backward} pass to finish
before it starts averaging. It packs gradients into \(\sim\)25\,MB \emph{buckets} as
\textit{backward} walks from the last layer to the first~\citep{li2020ddp} and launches each
bucket's all-reduce the moment it is full. If the network engine can run without stalling
the GPU, early buckets finish while the GPU is still working on earlier layers, and only
the last bucket is \emph{exposed communication}. We choose Ray to decide \emph{which} Macs join a
job; the process group decides \emph{how} they exchange tensors once it starts.

\begin{table}[!b]
\centering
\footnotesize
\caption{How distributed backends handle gradients across accelerators.}
\label{tab:backends}
\begin{tabularx}{\columnwidth}{@{}p{0.21\columnwidth}p{0.19\columnwidth}X@{}}
\toprule
Backend & Device & How gradients get averaged \\
\midrule
NCCL / RCCL & CUDA / ROCm & GPU kernels \& DMA over NVLink, RDMA, or TCP \\
XCCL / HCCL & XPU / NPU / HPU & vendor \code{c10d::Backend} plugins (Intel, Ascend, Gaudi) \\
Gloo / MPI & CPU RAM & CPU-only; stock PyTorch fails on MPS models, and manual staging stalls the GPU \\
MLX & \code{mx.array} & separate framework; Ethernet TCP/MPI paths are CPU-bound, not PyTorch MPS \\
\mccl{} (Ours) & Apple MPS & averages in place in unified memory; Metal shared events order GPU and CPU \\
\bottomrule
\end{tabularx}
\end{table}

\begin{figure*}[t]
\centering
\begin{tikzpicture}[x=1cm,y=1cm,
  lane/.style={font=\footnotesize\bfseries, anchor=east},
  k/.style={draw, rounded corners=1.5pt, minimum height=0.45cm,
            font=\scriptsize, inner sep=2pt},
  ev/.style={-{Stealth[length=1.6mm]}, thick, red!65!black},
  lbl/.style={font=\scriptsize, fill=white, inner sep=1pt}]
  \draw[gray!50] (0,2) -- (12.8,2);
  \draw[gray!50] (0,0) -- (12.8,0);
  \node[lane] at (-0.1,2) {GPU stream};
  \node[lane] at (-0.1,0) {CPU engine};
  \node[k, fill=blue!10, minimum width=2.3cm] (bw1) at (1.25,2)
       {\textit{backward} writes \(b_k\)};
  \node[k, fill=red!12, minimum width=0.9cm] (sig) at (3.0,2) {signal \(k\)};
  \node[k, fill=blue!10, minimum width=3.3cm] (bw2) at (5.3,2)
       {rest of \textit{backward}};
  \node[k, fill=red!12, minimum width=0.9cm] (wt) at (7.6,2) {wait \(k\)};
  \node[k, fill=blue!10, minimum width=2.6cm] (opt) at (9.55,2)
       {optimizer reads \(b_k\)};
  \node[k, fill=orange!15, minimum width=3.9cm] (red) at (5.1,0)
       {reduce + send/receive \(b_k\) in place};
  \node[k, fill=red!12, minimum width=0.9cm] (sig2) at (7.6,0) {signal \(k\)};
  \draw[ev] (sig.south) to[out=-90,in=150] node[lbl,pos=0.45,left]{O1}
            (red.north west);
  \draw[ev] (sig2.north) to[out=90,in=-90] node[lbl,pos=0.5,right]{O2}
            (wt.south);
  \draw[gray!75, dashed, thick] (2.55,2.55) -- (8.05,2.55);
  \draw[gray!75, thick] (2.55,2.45) -- (2.55,2.65)
                        (8.05,2.45) -- (8.05,2.65);
  \node[font=\scriptsize, text=gray!75] at (5.3,2.78)
       {O3: PyTorch thread is the sole stream committer};
\end{tikzpicture}
\caption{How one gradient bucket \(b_k\) changes hands without stopping the GPU. A
ticket signal queued right after the kernels that write \(b_k\) tells the CPU it may
start (O1); a second ticket raised when the ring finishes gates the optimizer's read
(O2); only PyTorch's thread ever commits the GPU stream (O3).}
\label{fig:protocol}
\end{figure*}
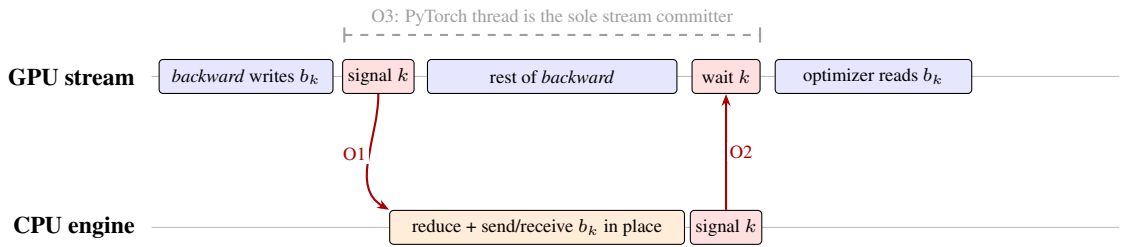

\paragraph{Macs are left out.}
CUDA, ROCm, Intel, Ascend, and Gaudi devices all join PyTorch through
the same \code{c10d::Backend} interface~\citep{ptcustombackends}
(\cref{tab:backends}), but Apple's GPU has no such backend. On macOS,
stock PyTorch offers only Gloo, which operates on CPU memory. Wrapping
an MPS model in \code{DistributedDataParallel} with Gloo fails before
the first training step completes, and staging each bucket to the CPU
by hand works only by draining the MPS stream for every bucket.
Apple's MLX provides its own collectives, but as a separate framework
it does not interoperate with PyTorch's MPS stream or caching
allocator. MLX's Ethernet and MPI paths reduce on the CPU, while its fastest path, Thunderbolt RDMA, requires a direct link between every pair of Macs, which limits a fully connected cluster to about six Macs.
\mccl{} also supports Thunderbolt, but we target switched Ethernet because Thunderbolt requires direct cabling and our workstations span dozens of rooms.

\paragraph{Unified memory, shared events, and wire rate.}
On an M4 Max, the CPU and GPU share the same 128\,GB of DRAM, with no
PCIe bus between them, so a CPU thread can average a GPU gradient in
place. Sharing memory costs little bandwidth. On single-die Apple silicon, the
CPU and GPU each reach about 90\% of the advertised unified-memory
bandwidth~\citep{beischl2026apple,hubner2025apple}. What
sharing does require is explicit ordering. Atomic operations are not
coherent across the two devices~\citep{beischl2026apple}, and PyTorch
queues GPU kernels asynchronously onto a single MPS command stream, so
a CPU thread cannot tell from memory alone whether the GPU has finished
writing a buffer. Metal shared events (\code{MTLSharedEvent}) supply that signal with a
counter that the GPU stream and CPU threads can both raise and wait on
without draining the device~\citep{applesharedevent}.

The network sets the other limit. 10\,Gb Ethernet carries 1.25\,GB/s
at line rate; with 9000-byte jumbo frames, a TCP socket delivers at
most 1.238\,GB/s of tensor payload, which we call \emph{TCP goodput}.
Because a ring all-reduce sends each byte around the ring twice
(\(2(N{-}1)/N\) wire bytes per tensor byte), we report \emph{bus
bandwidth}~\citep{nccltests}, which compares directly against TCP
goodput at any cluster size. Small messages are limited instead by
latency, chiefly the 0.59\,ms switch round trip.
\section{Cluster Setup and Software Stack}
\label{sec:setup}

Our controlled testbed (\cref{fig:system}) is a 24-node slice of
Skywalker Sound's workstation fleet. Each node is an Apple M4 Max Mac
Studio (16 CPU cores, 40-core GPU, 128\,GB unified memory) on a single
10\,Gb Ethernet switch, configured with 9000-byte jumbo frames and
32\,MiB socket buffers. The nodes run PyTorch 2.14.0 and \mccl{} 6.8.0
(commit \code{64a359b}) on macOS 26.4 or 26.5.1.

We launch jobs with Ray Train~\citep{moritz2018ray}, the same launcher
we use on our NVIDIA clusters. A Linux head node runs the Ray 2.56.1
controller and places PyTorch Lightning 2.1.0
workers~\citep{falcon2019lightning} with a \code{SPREAD} placement
group, so each worker lands on a different Mac and a job across dozens
of machines starts with one command rather than a \code{torchrun}
invocation on every Mac. The job script is identical to the one we run
on GPUs, apart from the head node it connects to.

Neither framework supports this setup as shipped, and each refuses one
thing. Ray discovers GPUs through CUDA, so every Mac reports zero
GPUs~\citep{ray39136}. We instead give each Mac one unit of a custom
resource (\code{pool:mac}) and have every worker request it, which
both makes Macs schedulable and caps each machine at one worker, so
two workers never compete for the same GPU and memory. Lightning
treats MPS as a single-device accelerator and rejects parallel
strategies on it~\citep{lightningmps,ray49858}, and its stock DDP
strategy passes CUDA \code{device_ids}, which the MPS process group
does not accept. A small custom strategy pins each worker to the \code{mps} device, returns \code{device_ids=None}, and drops the NCCL environment
variables Ray sets for GPU jobs. With these changes, importing \code{mccl} registers the \code{mccl} backend for MPS tensors, and a training job on the Mac fleet runs the same way it does on our CUDA clusters, with the same Ray
Train script, the same Lightning module, and
\code{DistributedDataParallel} with \code{mccl} in place of NCCL.

\section{MCCL Design: The Shared-Memory Handoff}
\label{sec:design}

Halfway through a \textit{backward} pass, the GPU finishes computing a 25\,MB gradient bucket for
the later layers of the model while continuing to compute gradients for earlier layers.
\mccl{} must average that bucket across 24 Macs on CPU threads before the optimizer step, under two constraints. The CPU cannot
read the bucket until the GPU has finished writing it, and the
optimizer cannot read it until the network has finished averaging it.

\subsection{In-Place Reduction in Unified Memory}
\label{sec:uma}

When PyTorch allocates an MPS tensor in shared Metal
storage~\citep{applemetalstorage}, the GPU's \code{MTLBuffer} and the
CPU's pointer are the same DRAM pages, and since PyTorch 2.12 every MPS
tensor is allocated this way~\citep{pt175818}. \mccl{} works on those bytes directly. It takes the CPU pointer to the
gradient bucket,
streams slices out over TCP, and uses Apple Accelerate (\code{vDSP}) to
add incoming slices from peer Macs into the bucket, accumulating fp16
and bf16 in fp32. vDSP runs on the CPU's SIMD units, and its
element-wise adds are far faster than the 1.238\,GB/s the network
delivers, so the reduction never limits the ring. When the ring
finishes, the averaged gradients are already where the optimizer
expects them.

We reduce in place not to save memory bandwidth, which the M4 Max has
in abundance, but to keep the GPU running. Copying a bucket to a
separate CPU buffer would require draining the GPU command stream for
every bucket; working in place with event tickets lets \textit{backward}
continue. Before PyTorch 2.12, tensors could live in device-only memory
that the CPU cannot address, and \mccl{} must take that slower path,
blitting each bucket through a shared staging buffer and waiting for
the copy. The ablation in \cref{sec:eval-wire} forces the same path and
shows what it costs.

\begin{figure*}[t]
\centering
\begin{tabular}{@{}c@{\hspace{0.02\textwidth}}c@{}}
\maybefig{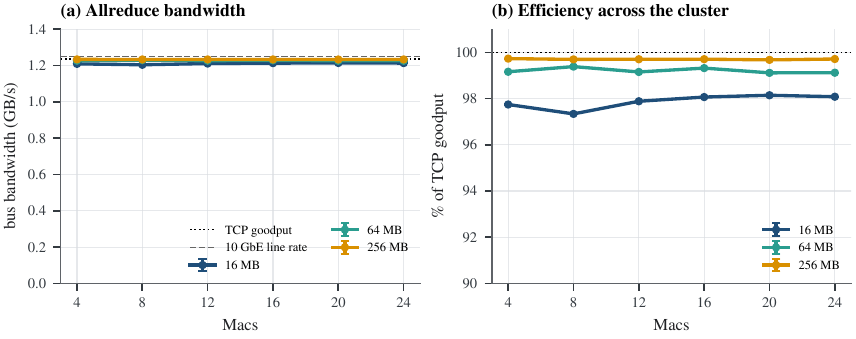}{0.64\textwidth} &
\maybefig{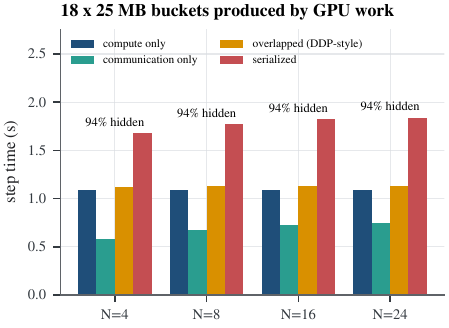}{0.30\textwidth}
\end{tabular}
\caption{Scaling from 4 to 24 Macs. Left and center: for training-sized buckets
(16--256\,MB), all-reduce bus bandwidth stays flat at 98--99.7\% of the 10\,GbE TCP
ceiling through 24 nodes. Right: in a bucketed training step, \mccl{} hides 94\% of
communication behind GPU \textit{backward}; blitting through a Metal staging pool and waiting for
completion (\code{MCCL_UNIFIED_COLLECTIVE=0}) eliminates that overlap.}
\label{fig:performance}
\end{figure*}

\noindent\begin{minipage}{\columnwidth}
\subsection{Per-Bucket Shared-Event Ordering}
\label{sec:sync}

When PyTorch hands \mccl{} a bucket pointer, the GPU usually has not written the bucket
yet. To keep the CPU and GPU in step without stalling either one, \mccl{} enforces three
rules (\cref{fig:protocol}):
\begin{enumerate}
  \item[\textbf{O1}] \emph{Producer ordering:} the CPU engine does not read bucket
        \(b_k\) until the GPU kernels that write \(b_k\) have completed.
  \item[\textbf{O2}] \emph{Consumer ordering:} the GPU's optimizer kernel does not read
        \(b_k\) until the CPU engine has written the final reduced slice into memory.
  \item[\textbf{O3}] \emph{Single-thread stream ownership:} only PyTorch's main thread
        encodes or commits commands on PyTorch's MPS stream.
\end{enumerate}
\end{minipage}
\vspace{3pt}

O1 and O2 are enforced by two ticket counters backed by Metal's
\code{MTLSharedEvent}~\citep{applesharedevent}, and bucket~$k$ passes
through three hands. PyTorch's thread queues the kernels that write
it, encodes a signal on producer ticket~$k$, flushes the command
buffer, and moves on to bucket~$k{+}1$ without waiting. When the GPU
finishes those kernels, the ticket increments, and \mccl{}'s CPU
thread, which has been waiting on it, sends bucket~$k$ into the TCP
ring. When the ring completes, the CPU thread raises consumer
ticket~$k$. When PyTorch later calls \code{wait()} before the
optimizer step, it encodes a GPU-side wait on that ticket ahead of the
optimizer kernels, so nothing blocks on the CPU. \mccl{} holds
bucket~$k$ until \code{wait()} returns, so PyTorch's caching allocator
cannot reuse its \code{MTLBuffer} while the CPU is still writing the
reduced slice.

We considered two simpler designs. Calling
\code{torch.mps.synchronize()} before each bucket satisfies O1 and O2,
but it drains the GPU 18 times per ViT step, removing the overlap that
bucketing is meant to provide. Committing PyTorch's MPS stream from
\mccl{}'s background thread avoids that stall but violates O3, and in
our tests it crashed inside Metal because PyTorch's stream wrapper is
not thread-safe. PyTorch's public \code{torch.mps.Event} covers only half of this
ordering. A CPU thread can wait on an event recorded by the GPU stream,
which satisfies O1, but as of PyTorch 2.14, which we use, no
public API lets a CPU thread signal an event that GPU work then waits
on, which O2 requires. PyTorch 2.15 adds public MPS streams and
\code{Tensor.record_stream}~\citep{pt215}; these may simplify parts of
\mccl{}'s integration with PyTorch's MPS stream, such as buffer
lifetime tracking, and we plan to evaluate them once 2.15 is stable.

\subsection{Pipelined TCP Ring and Tree Transport}
\label{sec:transport}

Once the CPU thread owns a bucket, \mccl{} uses conventional
algorithms, sending buckets above 256\,KiB around a pipelined
ring~\citep{patarasuk2009bandwidth} and smaller messages through a
$\log_2 N$ recursive-doubling tree~\citep{thakur2005mpich}. The harder problem is that Macs finish \textit{backward} at different times. \mccl{} has no
per-bucket barrier to line them up, and a ring sender is gated only by
its own receives, so a rank that finishes \textit{backward} a few milliseconds
early can run almost the entire reduce-scatter before its slower
neighbor has consumed anything. Left unchecked, that rank would push a
whole DDP bucket into the receiver.

\paragraph{Parking and credits.} A dedicated reader thread per peer
lets the slower rank accept data before it is ready to use it. The
reader demultiplexes each incoming frame by its operation, sequence
number, and tensor ID and parks it until the receiving rank posts the
matching step, up to a fixed park limit. To bound that buffer, the
receiver sends a 1-byte credit back to the sender after consuming each
chunk of 1\,MB or more, which caps the sender's lead at the pipeline
depth plus two chunks. The park buffer therefore holds a few chunks
rather than a whole bucket, and a fast Mac runs ahead by that bounded
amount instead of waiting for its neighbors. The same mechanism makes a per-collective barrier unnecessary. The
TCPStore barrier it replaces would cost one set plus 23 serial round
trips on rank~0 for every collective at 24 nodes.

\paragraph{Issue order.} Collectives must also enter the ring in the
same order on every rank. In an early version with two pool threads, a
plain mutex let one Mac start bucket~$k{+}1$ while its neighbor was
still in bucket~$k$, and each then waited indefinitely for the other's
traffic (\cref{sec:races}). \mccl{} now assigns each collective a
ticket on the thread that called it, PyTorch's main thread, and admits
collectives to the wire strictly in ticket order.

\section{Controlled Evaluation}
\label{sec:eval}

We evaluate three aspects in order. The first is whether the ring
saturates the 10\,Gb link, the second is whether that traffic stays
hidden behind \textit{backward} computation, and the third is whether training
converges correctly at every cluster size. We use microbenchmarks to answer the
first two. For the third, we train on CIFAR-10 rather than on our audio
data, because a communication backend is indifferent to what the
gradients represent. Its behavior depends only on the bytes exchanged
per step and the \textit{backward} time available to overlap them. CIFAR-10 is
public and inexpensive, so others can rerun these experiments on their
own Macs and use the results to estimate how a model with similar
gradient volume, such as an audio VAE, a stem separator, or a diffusion
refiner, would scale. Here, we use two models. A 113.5M-parameter Vision
Transformer is the throughput and communication stress case, and its
454\,MB of gradients per step is roughly three times what our audio VAE
exchanges. A 31.6M-parameter GroupNorm ResNet tests convergence across
cluster sizes. We close with correctness, shared storage, and sustained
reliability.

\subsection{Wire Scaling and Compute Overlap}
\label{sec:eval-wire}

We first checked whether the 10\,Gb network limits all-reduce
throughput as the cluster grows. In a ring, each Mac exchanges data
only with its two neighbors, so adding Macs adds hops but not per-link
traffic. Consistent with this, 16--256\,MB buckets move at
1.21--1.23\,GB/s of bus bandwidth from 4 to 24 Macs, 98--99.7\% of the
1.238\,GB/s that jumbo-frame TCP can carry on a 10\,Gb link
(\cref{fig:performance}, left and center). Small messages behave
differently. A 1\,MB all-reduce across 24 Macs puts only 42\,KB on each
hop, so the 0.59\,ms switch round trip dominates, which is why \mccl{}
sends messages under 256\,KiB over a recursive-doubling tree.

We next measured how much of that traffic is hidden behind \textit{backward}
computation, using a synthetic step with eighteen 25\,MB buckets timed
four ways, as compute alone, communication alone, overlapped, and
serialized (\cref{fig:performance}, right). From 4 to 24 Macs, about
94\% of communication time is overlapped with \textit{backward}. At 24 Macs,
0.74\,s of all-reduce traffic over 1.09\,s of GPU work adds 45\,ms to
the step, a 4.1\% overhead. When we disable the in-place path
(\code{MCCL_UNIFIED_COLLECTIVE=0}), each bucket is copied through a
Metal staging buffer and waited on, the same path a tensor in
device-only memory would take before PyTorch 2.12, and overlap drops to
zero.

\subsection{Training Throughput and Convergence}
\label{sec:throughput}

\begin{figure*}[t]
\centering
\begin{minipage}[t]{0.49\textwidth}
\vspace{0pt}
\centering
\maybefig{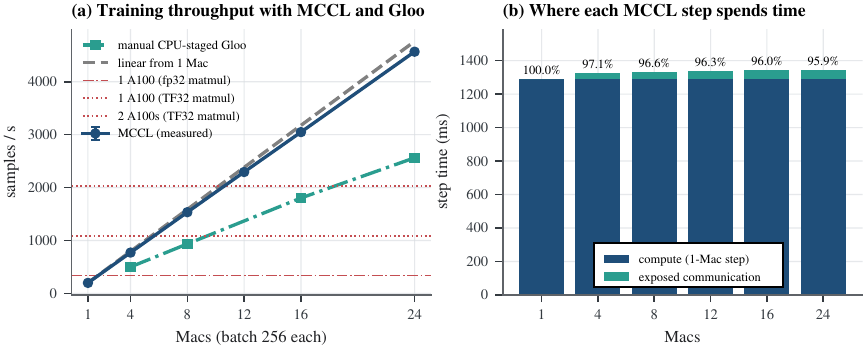}{\textwidth}
\caption{Vision Transformer throughput (113.5M parameters, batch 256 per device,
454\,MB of fp32 gradients per step) from 1 to 24 Mac Studios, against manual CPU-staged
Gloo and NVIDIA A100 80\,GB GPUs with TF32 or fp32 matmul.}
\label{fig:cifar-main}
\end{minipage}\hfill
\begin{minipage}[t]{0.49\textwidth}
\vspace{0pt}
\centering
\maybefig{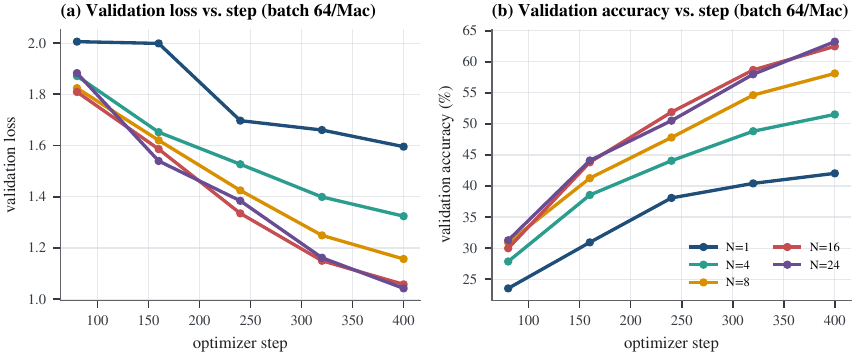}{\textwidth}
\caption{Validation loss and accuracy vs.\ optimizer step for a
31.6M-parameter GroupNorm ResNet on CIFAR-10, from 1 to 24 Macs at
batch 64 per Mac (global batch 64 to 1,536), with the AdamW learning
rate scaled as $4\times10^{-4}\cdot K^{0.45}$ for $K$ Macs. All
cluster sizes converge stably, and larger clusters make more progress
per step because each step averages over more examples.}
\label{fig:conv-main}
\end{minipage}
\end{figure*}

A custom Vision Transformer is our stress test, with 113.5M parameters, batch 256 per Mac, and eighteen 25\,MB buckets per step. Seventeen of those buckets
finish averaging while the GPU is still working on earlier layers, so the optimizer only
ever waits for the last one. Going from 1 to 24 Macs adds 55\,ms to a 1,290\,ms step.
Throughput reaches \textbf{4,568\,samples/s}, \textbf{95.9\%} of linear.

The strongest alternative in stock PyTorch is a manual Gloo reducer,
which synchronizes the MPS stream after \textit{backward}, copies each bucket to
the CPU, and reduces it with Gloo. That path takes 2,398\,ms per
step at 24 Macs (2,562\,samples/s), so \mccl{} is \textbf{\(1.78\times\) faster} on the
same hardware. To benchmark against our on-prem NVIDIA GPU clusters, we ran the same script on NVIDIA A100
80\,GB GPUs at batch~256 per device. With TF32 matmul (\texttt{high}), one A100 reaches
\textbf{1,088\,samples/s} (about \textbf{5.5 Macs}) and two reach \textbf{2,028}, so 24
Macs deliver \textbf{\(2.25\times\)} a two-A100 server. With fp32 matmul
(\texttt{highest}, matching the Macs), one A100 drops to \textbf{340\,samples/s}, about
\textbf{\(1.7\times\)} one Mac. Turning on \code{bf16} would help the A100s more than the
Macs, where an \(8192^2\) GEMM only moves from 13.4 to 15.1\,TFLOP/s between fp32 and
bf16.

Throughput is useful only if training still converges, and a
113.5M-parameter ViT overfits $32{\times}32$ CIFAR-10 at large global
batches, so we test convergence with a 31.6M-parameter ResNet
(\cref{fig:conv-main}). GroupNorm keeps normalization independent of
per-rank batch size, and the model is a lighter communication case than
the ViT, with five 25\,MB buckets rather than eighteen and a shorter
\textit{backward} pass to hide them behind. Every run takes 400 optimizer steps
at batch 64 per Mac, with the AdamW learning rate scaled as
$4\times10^{-4}\cdot K^{0.45}$ for $K$ Macs, so the global batch grows
from 64 at one Mac to 1,536 at 24. All runs converge without divergence
or instability. After 400 steps, validation loss falls from 1.60 at one
Mac to 1.04 at 24, and test accuracy rises from 42.0\% to 63.2\%,
consistent with the larger global batch at each step.

\subsection{Studio Storage and Sustained Reliability}
\label{sec:storage}

Our audio data lives on central ZFS servers that the Macs mount over
SMB or NFS. NFS lists directories faster on a cold start (0.9 vs.\
10.6\,s for thousands of files), while SMB streams data faster once
loading begins (757 vs.\ 538\,MB/s per Mac). Precomputing the file list into a manifest removes
the cold SMB walk, and after that neither protocol changes step time. To isolate
transport correctness under sustained load, we ran a 10-minute stress test across all 24 Macs. It completed 17,784
back-to-back all-reduces per rank, 0.34\,TB of traffic per Mac, with no
bit errors, while \mccl{} used 0.31 of one CPU core. Our longest internal training runs span over 3 weeks with no deadlocks or training instability.

\section{Lessons from Scaling Past Two Nodes}
\label{sec:issues}
\label {sec:races}

\begin{figure*}[t]
\centering
\begin{minipage}[t]{0.50\textwidth}
\vspace{0pt}
\centering
\maybefig{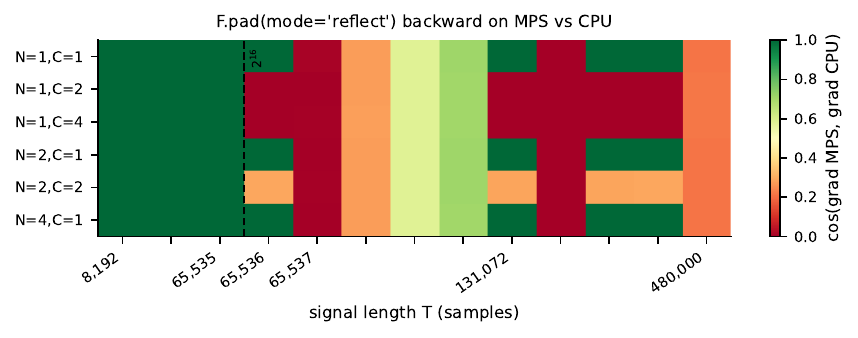}{\textwidth}
\caption{Cosine similarity between MPS and CPU \code{float64} reference gradients for
\code{F.pad(..., mode="reflect")}. \textit{Forward} outputs are exact everywhere; on MPS, the
\textit{backward} pass is exact up to 65,535 samples (\(2^{16}{-}1\)), fails above \(2^{16}\) for
\(N{=}1, C{=}1\) except at exact multiples of \(2^{16}\), and fails at every length
\(\ge 2^{16}\) whenever \(N{>}1\) or \(C{>}1\).}
\label{fig:padmap}
\end{minipage}\hfill
\begin{minipage}[t]{0.47\textwidth}
\vspace{0pt}
\centering
\footnotesize
\captionof{table}{Centered STFT \textit{backward} accuracy and 300-step waveform-recovery loss
(initial loss 1.821) for signals with \(T \ge 2^{16}\) samples (batch 2, window 2048,
hop 512).}
\label{tab:stft-main}
\vspace{4pt}
\setlength{\tabcolsep}{2.4pt}
\begin{tabular}{@{}lrrr@{}}
\toprule
STFT backend (MPS) & Min $\cos(\nabla)$ & Final loss & Fwd+bwd \\
\midrule
PyTorch \code{torch.stft} & $-0.002$ & 5.525 & 1.1\,ms \\
\code{mps-spectro} (default)$^\dagger$ & $-0.002$ & 5.525 & 0.8\,ms \\
\code{mps-spectro} (Metal) & \textbf{1.000} & \textbf{0.924} & \textbf{0.8\,ms} \\
\mccl{} Accelerate / \code{vDSP} & 0.997 & 0.921 & 13.7\,ms \\
\midrule
CPU \code{float64} reference & 1.000 & 0.924 & 14.7\,ms \\
\bottomrule
\multicolumn{4}{@{}l@{}}{\scriptsize $^\dagger$Default shape rule falls back to \code{torch.stft}.}
\end{tabular}
\end{minipage}
\end{figure*}


Our early development used two nodes and mainly ran smoke tests of
gradient correctness on synthetic data. Scaling to 24 nodes with real
datasets exposed three problems that did not appear with two nodes. We describe
how we found each one and what we changed as others building
distributed training on Apple silicon are likely to face the same
issues.

\subsection{Unified-Memory Races and Ring Ordering}

\paragraph{Buffer-reuse race in unified memory.}
At 24 nodes, we observed occasional corrupted gradients that disappeared
whenever we added a \code{print()} to investigate. The cause was a
temporary GPU tensor that an early version of \mccl{} allocated, copied
out of, and released when the call returned. Under load, PyTorch's
caching allocator handed the same \code{MTLBuffer} to the next
collective while the earlier copy was still pending in the command
queue, and \mccl{}'s CPU thread wrote incoming network bytes over a
buffer the GPU had not finished reading. Our unit tests never caught
it, because \code{tensor.item()} and \code{print()} both drain the MPS
stream between collectives. \mccl{} now receives directly into the
caller's output tensor and never allocates temporary GPU buffers on its
background thread, and it holds every bucket until \code{wait()}
returns (\cref{sec:sync}). We now also test asynchronous bursts with
every host synchronization removed, and all 640 operations per rank
across 24 Macs are exact.

\paragraph{Cross-rank bucket ordering in the ring.} At 24 nodes, the
ring occasionally deadlocked. PyTorch can finish two buckets a few
microseconds apart, and early versions of \mccl{} let worker threads
pick up ready buckets in whatever order the OS scheduled them. One rank
would sometimes start bucket~2 while its neighbor started bucket~1, and
each waited on the other. \mccl{} now stamps every collective with a
sequence number on PyTorch's main thread and admits work to the
transport only in that order.

\subsection{A Silent MPS Gradient Bug Above $2^{16}$ Samples}
\label{sec:padbug}

The most time-consuming problem we encountered was outside \mccl{}.
Our audio models train with multi-resolution centered short-time
Fourier transform (STFT) losses~\citep{yamamoto2020pwg}, and models
trained on short clips converged normally on MPS. When we moved our VAE
to 1.6\,s crops (76,800 samples at 48\,kHz), training stopped
improving, with no errors or NaN values to indicate a problem. Because
the same code trained correctly on our NVIDIA clusters, we compared
each layer's \textit{backward} pass on MPS against a CPU \code{float64}
reference.

The comparison pinned the fault on the MPS \textit{backward} kernel for
\code{F.pad(..., mode="reflect")}, which every centered
\code{torch.stft} calls (\cref{fig:padmap}). The \textit{forward} pass is
bit-exact, but the \textit{backward} pass is exact only up to 65,535 samples
($2^{16}{-}1$). Above that length, a single mono signal ($N{=}1$,
$C{=}1$) receives wrong gradients at every length except exact
multiples of $2^{16}$, and any input with more than one batch element
or channel fails at every length of $2^{16}$ or more, a pattern
consistent with an indexing overflow at $2^{16}$ elements.

\Cref{tab:stft-main} shows the effect on training. With native MPS
gradients, and with the default \code{mps-spectro} path, which hands
these shapes back to \code{torch.stft}, a 300-step waveform-recovery
loss rises from 1.821 to 5.525 instead of falling to 0.924. Forcing the
custom Metal kernels in \code{mps-spectro}~\citep{mpsspectro} restores
exact gradients at 0.8\,ms per call, and \mccl{} ships an independent
Apple Accelerate STFT as a permanent reference check. The bug affected
every PyTorch version we used, from 2.11 through 2.14, because PyTorch
delegated this kernel to Apple's MPSGraph~\citep{pt196949}. This was only recently fixed upstream by moving reflection padding to native Metal
kernels~\citep{pt196985}.

We now run every new model through this check before training it
across the cluster. It executes each layer's \textit{forward} and \textit{backward} pass
on both MPS and CPU \code{float64} at production tensor shapes and
flags any output or gradient whose cosine similarity to the CPU
reference falls below 0.99 or whose relative error,
$\lVert x_{\mathrm{MPS}} - x_{\mathrm{CPU}} \rVert_2 /
\lVert x_{\mathrm{CPU}} \rVert_2$, exceeds $10^{-3}$.

\section{From 24 Macs to Hundreds}
\label{sec:scale}
\label{sec:scaling}
Our evaluation used up to 24 nodes, and the facility has more than 200.
Assuming 100 machines are idle each night, the fleet corresponds to
roughly 18 A100 GPUs with TF32 matmul, and with all 200 available,
roughly 35. Reaching that scale requires solving two problems that do
not appear at 24 nodes.

The first problem is hop count. A flat ring splits each bucket into $N$
pieces and sends them $2(N{-}1)$ hops, and each hop pays a fixed switch
cost regardless of message size, 306\,$\mu$s in our model and
consistent with the measured 0.59\,ms round trip. A
latency-plus-bandwidth model~\citep{hockney1994communication} fitted to
the 4--24 Mac data matches every measured bucket of 16\,MB or more
within 1.9\%. The model predicts that a standard 25\,MB bucket stays
bandwidth-bound up to roughly 70 Macs, beyond which a flat ring spends
more time waiting on switch hops than moving bytes. It also predicts
that raising PyTorch's \code{bucket_cap_mb} to 50\,MB moves the
crossover past 128 Macs and holds 1.23\,GB/s at 100 nodes. These values
extrapolate the 4--24 Mac measurements and are not measurements at 70
or 100 nodes. For the full fleet, the standard answer is a two-level
ring~\citep{jia2018highly,mikami2018massively}, which reduces within
groups of 10--16 Macs on the same switch and then across switches.

The second problem is that the workstations belong to artists first,
so the training pool must grow and shrink with the studio's schedule.
An empty room does not mean an idle machine, because some editors work
from home and connect to their workstations remotely. A daemon on each
Mac would therefore need to detect active use on the machine itself,
characterizing active Pro Tools sessions or a remote login, before joining the
Mac to Ray, and remove it as soon as either appears. We leave the
design of this availability detection to future work.

Recovery when a Mac leaves is a separate problem. Ray Train's elastic
training~\citep{raytrainelastic} lets a job specify a range of workers
rather than a fixed count. When a Mac leaves, the job restarts from its
latest checkpoint with the remaining workers, and when more Macs become
available, it resizes the worker group to include them, so a job loses
at most the work since its last checkpoint. We have not yet tested
elastic training with \mccl{}. Each Mac would also cache dataset shards
on its local SSD, so that 200 workers starting at once do not overload
a single ZFS volume. Finally, \mccl{} already implements the
\code{allgather} and \code{reduce_scatter} calls that sharded FSDP
training~\citep{zhao2023fsdp} needs, though every job in this paper
keeps a full model copy per Mac.
\section{Limitations}
\label{sec:limitations}

\mccl{} hides communication only when the \textit{backward} pass is long
relative to gradient volume. At batch 64 per Mac, the ResNet's
126\,MB of gradients cannot be fully hidden at 24 nodes, and per-step
time rises from 227\,ms to 305\,ms. Models with small per-device
batches therefore need larger batches, gradient accumulation, or
reduced-precision gradients to scale well on our network.

The network itself is the larger constraint. NVIDIA training systems
connect GPUs with NVLink and NVSwitch at up to 900\,GB/s within a node
and InfiniBand at up to 400\,Gb/s between
nodes~\citep{vanderouderaa2025}, and NCCL~\citep{nccl} moves data over
these links with RDMA rather than kernel TCP. Our Macs share a single
10\,Gb Ethernet switch, which delivers 1.238\,GB/s of TCP goodput per
link and costs 0.59\,ms per round trip. Large buckets still fill that
link, but every collective pays the round trip, so small or frequent
messages are latency-bound long before they are bandwidth-bound. Faster
links would require Thunderbolt network adapters, since the Mac
Studio's built-in port is 10\,GbE, and we have not evaluated
lower-latency Ethernet or switched RDMA.

Our measurements stop at 24 nodes on one switch, and the predictions in
\cref{sec:scaling} are extrapolations. \mccl{} also does not yet handle
preemption itself. A timeout or peer failure errors all in-flight
collectives, and recovery relies on restarting from the last
checkpoint.

Finally, we have not evaluated distributed inference, although \mccl{}
exposes the collectives it needs. Our measurements suggest how the
common strategies would fare. Replicated inference, in which each Mac
holds a full model copy and serves its own requests, needs no
per-request collectives and should scale with the number of Macs, and
128\,GB of unified memory fits large models on a single machine.
Pipeline-parallel inference sends one small activation per stage
boundary per token, so each boundary would add on the order of half a
switch round trip, about 0.3\,ms. Tensor-parallel inference is the
poor fit. It requires two all-reduces per transformer layer per token,
each paying at least one 0.59\,ms round trip regardless of message
size, so a 32-layer model would spend roughly 38\,ms per token on
communication alone, capping decoding near 26 tokens per second per
sequence. On this network, we would therefore start with replicated
and pipeline-parallel inference and measure both before trying tensor
parallelism.
\section{Conclusion }
\label{sec:conclusion}

\mccl{} lets us run the same PyTorch Lightning code on our Mac fleet
that we run on our NVIDIA clusters. On 24 Mac Studios over 10\,GbE, it
reaches 98--99.7\% of TCP goodput, overlaps about 94\% of communication
with \textit{backward} computation, and scales a 113.5M-parameter ViT at 95.9\%
efficiency. The main contribution is not a new collective algorithm,
but a way to reduce gradients directly in unified memory without
stalling the GPU. Per-bucket Metal shared events coordinate the CPU
network threads with PyTorch's asynchronous MPS stream.

Scaling to 24 nodes also exposed problems we had not encountered in
smaller tests, including a buffer-reuse race, a ring-ordering deadlock,
and a silent MPS gradient bug. We have tested \mccl{} on 24 of the
facility's more than 200 Macs. Using the full fleet will require
hierarchical collectives, local data caching, and a way to release
workstations when artists need them. These are the next problems we
plan to address.

\paragraph{Availability.} \mccl{} is open source at
\url{https://github.com/mps-ddp/mccl}.


\section*{Appendix: Related Work}
\label{sec:related}

\paragraph{Custom PyTorch backends and Apple-silicon ML.}
PyTorch's \code{c10d::Backend} plugin API~\citep{ptcustombackends} is how every
non-NVIDIA accelerator joins \code{torch.distributed}: RCCL for AMD~\citep{rccl}, oneCCL
for Intel XPUs~\citep{oneccl}, HCCL plugins for Ascend and Gaudi~\citep{ascendhccl,
habanahccl}, and UCC at HPC sites~\citep{torchucc}. None of them targets MPS, and the
CPU options on macOS (Gloo~\citep{gloo}, Horovod/MPI~\citep{sergeev2018horovod}) only
move host buffers. Apple's MLX~\citep{mlx2023,mlxjaccl} ships TCP \code{ring}, OpenMPI,
and Thunderbolt~5 RDMA (\code{jaccl}) collectives, but for its own graph rather than
PyTorch. Two recent PyTorch pull requests target MPS collectives over Thunderbolt~5
RDMA: \#181087 was closed without merging, and TCCL is an open proposal that needs
Thunderbolt~5 cables and macOS~26.2~\citep{pt181087,tccl2026}. \mccl{} instead targets
the Mac workstations and switched 10\,Gb Ethernet a studio already has.
Exo~\citep{exo2024} targets distributed inference rather than training.

\paragraph{Collective algorithms and overlap.}
Pipelined rings and recursive-doubling trees are
classic~\citep{patarasuk2009bandwidth,thakur2005mpich}; NCCL and follow-on work tune
them to specific fabrics and hierarchies~\citep{nccl,hu2025demystifying,shah2023taccl,
desensi2024swing,cho2019blueconnect}, PyTorch buckets gradients to overlap transfers
with \textit{backward}~\citep{li2020ddp}, and ZeRO/FSDP shard model
state~\citep{rajbhandari2020zero,zhao2023fsdp}. \mccl{} does not add a new ring
algorithm. Its contribution is the unified-memory handoff that lets a CPU TCP engine
average MPS gradient buckets in place while PyTorch's GPU stream keeps running.

\bibliographystyle{mlsys2025}
\bibliography{refs}
\end{document}